\documentclass[preprint,aps,eqsecnum,nofootinbib,amsmath,amssymb,tightenlines]{revtex4}
\usepackage{bm}

\begin{document}

\title{Hydrodynamic Fluctuations in Relativistic Superfluids}

\author{Manuel A. Valle}
\email{manuel.valle@ehu.es}
\affiliation
    {%
    Departamento de F\'\i sica Te\'orica, 
    Universidad del Pa\'\i s Vasco, 
    Apartado 644, E-48080 Bilbao, Spain
    }%

\date{\today}

\begin{abstract}
The Hamiltonian formulation of superfluids 
based on noncanonical Poisson brackets is studied in detail. The assumption 
that the momentum density is proportional to the flow of the conserved energy 
is shown to lead to the covariant relativistic theory previously suggested by 
Khalatnikov, Lebedev and Carter, and some potentials  in this theory are given explicitly. 
We discuss hydrodynamic fluctuations in the presence of dissipative effects and we derive  
the corresponding set of hydrodynamic correlation functions. Kubo relations for the 
transport coefficients are obtained.
\end{abstract}

\keywords{}

\maketitle


\section{Introduction}

Recently there has been considerable interest in the possible occurrence of 
superfluidity  in relativistic systems such as neutron stars or quark matter 
at very high densities~\cite{Alford:2001dt}. A common distinctive feature 
of these systems is the spontaneous breaking of (at least) 
a $U(1)$ symmetry associated with particle number.

At nonzero temperature, the dynamic description of a given system 
at large distance and time scales 
is based on hydrodynamic equations for the hydrodynamic modes.  
Generically these 
consist of quantities whose long-wavelenght fluctuations have a 
large lifetime which becomes infinite as $k\rightarrow 0$. The 
hydrodynamic variables include densities of conserved quantities 
and Goldstone modes of broken symmetries. 

A relativistic extension of
superfluid hydrodynamics of ${}^{4}$He~\cite{Landau,Hohenberg}
was previously suggested by Khalatnikov, Lebedev and Carter in
Refs.~\cite{Khalatnikov,Khalatnikov2,Carter}. 
More recently, Son outlined the construction of the 
nondissipative hydrodynamics of relativistic systems 
with broken symmetries using the Poisson bracket method 
in the case of broken $U(1)$ symmetry~\cite{Son1} and in 
the case of nuclear matter~\cite{Son2}, where the chiral symmetry $SU(N_{f})_L \times SU(N_{f})_{R}$
is approximately broken down to $SU(N_{f})_{L+R}$. 

In this article we will pursue the scarce research on this subject.
In particular, in Sec.~\ref{sec:bra}
we derive in detail the equations of motion 
of superfluids using the Poisson bracket method, and we show 
that the assumption that 
the momentum density is proportional to the flow of the conserved energy 
leads to relativistic hydrodynamics of Khalatnikov and Lebedev. 
In this framework, with the alternative assumption that the momentum density 
is proportional to the flow of the $U(1)$ conserved charge one obtains 
the two-fluid model of non-relativistic superfluidity. 
In Sec.~\ref{sec:fluc} we quickly review the memory function formalism to study the 
hydrodynamic fluctuations and examine some sum rules, following closely the treatment given in 
Ref.~\cite{Forster}. 
We use this information, together the information about the forces and the memory matrix, to deduce   
the correlation funcions, the Kubo relations and the linearized constitutive relations in 
Sec.~\ref{sec:kubo}. As well, some Ward identities are checked.  

Although perhaps none of the statements 
made below could go beyond what is known about 
the equations of motion of relativistic superfluids, 
it can be of some interest to present the discussion of 
the main results from the point of view of the theory of 
hydrodynamic fluctuations and correlation functions, in order to 
show how this general framework organizes the physical description in the  
low-energy limit.


\section{Poisson bracket and hydrodynamic equations}
\label{sec:bra}
We first consider the low energy, nondissipative dynamics of a system 
without 
long-range interactions 
with broken $U(1)$ symmetry at a temperature well below criticality.  
In such a system, 
the hydrodynamic variables are five conserved densities, namely,   
the entropy per unit volume $s$, the density $n$ of the 
$U(1)$ charge and the
momentum densities $g^i$, plus a Goldstone mode $\varphi$.  
The dynamics is governed by the Hamiltonian functional 
$H[s,n,g^i,\partial_j\varphi]$ whose specific form 
can be computed or guessed from the 
underlying microscopic model. Note that the 
invariance of $H$ under $U(1)$ transformations, $
\delta \varphi = \alpha$, prevents the dependence upon  
$\varphi$.  The energy functional is 
an extensive quantity 
\begin{equation}
    H = \int d^3 \mathbf{x}\, \varepsilon,
\end{equation} 
where $\varepsilon(s,n,g^i,\partial_j\varphi)$ is the
energy density from which one can compute thermodynamic equilibrium
properties.  As the following discussion concerns the hydrodynamic
behaviour for large wavelength and low frequency,
it will be enough 
to consider the dependence of $\varepsilon$ up to first derivatives 
of $\varphi$. 
In order to obtain all the equations of motion, we must 
specify a Poisson bracket structure $[F, G]_{\mathrm{PB}}$ 
between functionals of the hydrodynamic 
variables. Then, the time derivative of a functional $\mathcal{V}$ of 
$\left\{s,n,g^i,\varphi\right\}$ 
can be derived from 
\begin{equation}
      \frac{\partial \mathcal{V}}{\partial t} = 
      \left[\mathcal{V}, H\right]_{\mathrm{PB}}.
\end{equation}
The noncanonical Poisson bracket is taken to be
\begin{eqnarray}\label{eq:PB}
    \left[F, G \right]_{\mathrm{PB}} & = & -\int d^3 \mathbf{x}\left[ 
     g^i
     \left(\frac{\delta F}{\delta g^j}\frac{\partial}{\partial x^j}\frac{\delta G}{\delta g^i}- 
           \frac{\delta G}{\delta g^j}\frac{\partial}{\partial x^j}\frac{\delta F}{\delta g^i} 
	   \right)\right.  \nonumber \\
     & &{}+n 
     \left(\frac{\delta F}{\delta g^j}\frac{\partial}{\partial x^j}\frac{\delta G}{\delta n}- 
           \frac{\delta G}{\delta g^j}\frac{\partial}{\partial x^j}\frac{\delta F}{\delta n} \right) 
           \nonumber \\
     & &{}+s
     \left(\frac{\delta F}{\delta g^j}\frac{\partial}{\partial x^j}\frac{\delta G}{\delta s}- 
           \frac{\delta G}{\delta g^j}\frac{\partial}{\partial x^j}\frac{\delta F}{\delta s} 
	   \right) \nonumber \\ 
     & &{}- \left.
     \frac{\partial \varphi}{\partial x^j} 
     \left(\frac{\delta F}{\delta g^j}\frac{\delta G}{\delta \varphi}- 
           \frac{\delta G}{\delta g^j}\frac{\delta F}{\delta \varphi }\right) 
     -
       \left(\frac{\delta F}{\delta n} \frac{\delta G}{\delta \varphi} - 
             \frac{\delta G}{\delta n} \frac{\delta F}{\delta \varphi} \right)  \right].
\end{eqnarray}
The form of the bracket in the first three terms follows from 
the conservation of the linear momentum, the $U(1)$-charge and the 
entropy~\cite{Morrison}. 
The remainder terms reflect the transformation property of 
the $\varphi$-field under
infinitesimal spatial translations and the fact
that the charge density and the Goldstone 
mode are canonically conjugated~\cite{Son1}
\begin{equation}\label{eq:com}
 \left[n(\mathbf{x}),\varphi(\mathbf{y})\right]_{\mathrm{PB}} = 
 \delta(\mathbf{x}-\mathbf{y}) .
\end{equation}
With this normalization the Goldstone mode $\varphi$ is dimensionless.

In terms of the quantities conjugate to the hydrodynamic variables, 
the temperature $T$, the
chemical potential $\mu$, the fluid velocity $\mathbf{v}$ and 
the vectorial quantity 
$\bm{\lambda}= \partial \varepsilon/\partial(\bm{\nabla}\varphi)$,
the energy density $\varepsilon$ satisfies the thermodynamic
relation 
\begin{equation}\label{eq:first}
 d \varepsilon = T d s + \mu d n  +
 \mathbf{v} \cdot d \mathbf{g} + 
 \bm{\lambda} \cdot d (\bm{\nabla}\varphi) .
\end{equation}
A second thermodynamic identity to be used below is the Gibbs-Duhem relation for the 
pressure, defined by 
\begin{equation}\label{eq:pressure}
p=-\varepsilon +T s + \mu n + \mathbf{v} \cdot \mathbf{g} .
\end{equation}
The non occurrence of $\bm{\lambda} \cdot  \bm{\nabla}\varphi$ in this expression
can be understood 
by noting that
the hydrodynamic Goldstone mode, which is not a conserved density,
cannot contribute to the total (integrated) entropy $S$ since it 
corresponds to a single coherent mode~\cite{Hohenberg}. 
Thus, the entropy is a homogeneous function 
of all extensive 
variables: the volume $V$ and the conserved quantities, 
energy, linear momentum and $U(1)$ charge. Accordingly,  
\begin{equation}
    S = \frac{\partial S}{\partial V} V + 
    \frac{\partial S}{\partial E} E +
    \sum_i \frac{\partial S}{\partial P^i} P^i + \frac{\partial S}{\partial Q} Q ,
\end{equation}
which, after multiplication by $V^{-1}$, yields the Gibbs-Duhem relation. 
Notice, however, that the thermodynamic identity for the pressure has the form 
\begin{equation}\label{eq:dp}
 d p  = s d T + n d \mu  +
 \mathbf{g} \cdot d \mathbf{v} - 
 \bm{\lambda} \cdot d (\bm{\nabla}\varphi) .
\end{equation}

From these formulas and the Poisson bracket it is easy to derive the equations of fluid
dynamics without dissipation.  
Let us list them.
The conservation laws adopt the form
\begin{eqnarray} 
&&{}\partial_{t} s(\mathbf{x}, t) + \bm{\nabla}\cdot (s \mathbf{v}) = 0, \\
&&{}\partial_{t} n(\mathbf{x}, t) + \bm{\nabla}\cdot (n \mathbf{v} + \bm{\lambda})  = 0 , \\ 
&&{}\partial_{t} g^i(\mathbf{x}, t) + \partial_{k} t^{k i} = 0 ,  
\end{eqnarray}
where the reactive part of the stress tensor is given by 
\begin{equation}
 t^{k i} = p\, \delta^{k i} + v^k g^i +\lambda^k \partial_{i}\varphi ,
\end{equation}
and the equation of motion for $\varphi$ is 
\begin{equation}
 \partial_{t} \varphi(\mathbf{x}, t) = -\mu 
 - \mathbf{v} \cdot \bm{\nabla} \varphi.
\end{equation}
As a consequence of these equations of motion and the Gibbs-Duhem 
relations~(\ref{eq:pressure}) and~(\ref{eq:dp}), we find the flow of the
conserved energy
\begin{eqnarray}
&&{}\partial_{t} \varepsilon(\mathbf{x}, t) + \partial_{i} j_{\varepsilon}^i = 0 , \\
\label{eq:je}
&&{}j_{\varepsilon}^i = (\mu + \mathbf{v} \cdot
\bm{\nabla}\varphi) \lambda^i+ (\varepsilon+p) v^i .
\end{eqnarray}
At this point, we can separate the momentum density $\mathbf{g}$ and
$\bm{\lambda}$ into two pieces proportional to $\mathbf{v}$ and
$\bm{\nabla}\varphi$ by\footnote{Because of
rotational invariance, this decomposition is completely general and
does not amount to a loss of generality.}
\begin{eqnarray}
\label{eq:ga}
\mathbf{g} &=& \alpha_{1}\mathbf{v} +
\alpha_{2}\bm{\nabla}\varphi ,\\
\label{eq:la}
\bm{\lambda} &=& \alpha_{3}\mathbf{v} + \xi \bm{\nabla}\varphi ,
\end{eqnarray}
where the $\alpha_{j}, \xi$ are considered as functions of
$(s,n,v^i,\partial_{j}\phi)$.
The quantity $\xi$ has dimension mass$^2$ and, as we shall see below,
characterizes the long wave limit of the equal-time correlator of
$\varphi$.  The symmetry of $t^{k i}$ requires that $\alpha_{3} =
-\alpha_{2}$.

Further progress can be made if we enforce Lorentz invariance of
the hydrodynamic equations.  This can be made by 
assuming that the momentum density is the same as the 
energy flow, $j_{\epsilon}^i = g^i$.  Consequently, from 
Eqs.~(\ref{eq:je}) and~(\ref{eq:ga}), we find 
\begin{eqnarray}
    \alpha_{1} &=& \varepsilon +p-\xi \left(\mu + \mathbf{v} \cdot
    \bm{\nabla}\varphi\right)^2, \\
    \alpha_{2} &=&\xi \left(\mu + \mathbf{v} \cdot \bm{\nabla}\varphi
    \right),
\end{eqnarray}
which lead to the constitutive relations 
\begin{eqnarray}
    t^{i j} &=& p\, \delta^{i j} + \left[\varepsilon +p-\xi \left(\mu +
    \mathbf{v} \cdot \bm{\nabla}\varphi\right)^2 \right] v^i v^j + \xi
    \partial_{i}\varphi \partial_{j}\varphi , \\
    \label{eq:gs}
    j_{\varepsilon}^i &=& \left[\varepsilon +p-\xi \left(\mu +
    \mathbf{v} \cdot \bm{\nabla}\varphi\right)^2 \right] v^i + \xi
    \left(\mu + \mathbf{v} \cdot \bm{\nabla}\varphi\right)
    \partial_{i}\varphi, \\
    \lambda^i &=& -\xi \left(\mu + \mathbf{v} \cdot
    \bm{\nabla}\varphi\right) v^i + \xi \partial_{i}\varphi .
\end{eqnarray}
It is important to emphasize that the energy flow is itself conserved because 
of the assumption of relativistic invariance. 
This fact will have implications on the form in which the
dissipative terms must be included in the full constitutive relations.

With the introduction of the four-vectors 
\begin{eqnarray}
&&{}\partial_\mu \varphi \equiv 
  (-\mu - \mathbf{v} \cdot \bm{\nabla}\varphi, \bm{\nabla}\varphi), \\ 
&&{} w_\mu \equiv (T + \mathbf{v} \cdot \mathbf{w},-\mathbf{w}), \\ 
&&{} J^\mu \equiv (n, n \mathbf{v} + \bm{\lambda}), \\ 
&&{} s^\mu \equiv (s, s \mathbf{v}),
\end{eqnarray}
where $\mathbf{w} \equiv  s^{-1} (\mathbf{g} - n \bm{\nabla}\varphi)$, it turns out that the 
differential Gibbs-Duhem identity can be written in a covariant form  as 
$dp = -J^\mu d(\partial_\mu \varphi) + s^\mu dw_\mu$, and the energy-momentum densities
and the stress tensor, $\varepsilon = T^{00}, j_\varepsilon^i = T^{0 i}, 
t^{i j} = T^{i j}$,  
are obtainable from the symmetric four-tensor
\begin{equation}
T^{\mu \nu} = -J^\mu \partial^\nu \varphi + s^\mu w^\nu  - p \eta^{\mu \nu} .
\end{equation}
Correspondingly, the conservation equations can be written as
\begin{eqnarray}
\partial_\mu T^{\mu \nu} &=& s^\mu (\partial_\mu w^\nu - \partial^\nu w_\mu) = 0, \\
\partial_\mu J^{\mu }&=& 0.
\end{eqnarray} 
These are in precise agreement with the relativistic formulation of
Lebedev, Khalatnikov and Carter~\cite{Khalatnikov,Khalatnikov2,Carter}
provided that $w^\mu$ has the explicit form\footnote{
I thank Dam T. Son for pointing out to me the 
overlap of these results with his unpublished work on the 
Poisson bracket approach to hydrodynamics~\cite{Son3}.} 
\begin{equation}
\mathbf{w} = \frac{1}{s}\left[\varepsilon +p-\xi \left(\mu +
\mathbf{v} \cdot \bm{\nabla}\varphi\right)^2 \right]\mathbf{v} -
\frac{1}{s}\left[n - \xi (\mu + \mathbf{v} \cdot \bm{\nabla}\varphi)
\right] \bm{\nabla}\varphi \;. 
\end{equation}
  
In this approach, it is very easy as well to derive the equations of the 
nonrelativistic two-fluid model.  It suffices to require the
proportionality between the momentum density $g^i$ and the flow of the
$U(1)$ conserved
charge, 
$g^i =  m (n v^i + \lambda^i)$, where $m$ is the mass of the only species of 
particle in the fluid. The consequences of this assumption in the context of 
effective Lagrangians for superconductors have been analyzed in
Ref.~\cite{Greiter}.  By introducing the superfluid density 
$n_s = m \xi$ and the superfluid velocity 
$v_s^i = m^{-1} \partial_i \varphi$, one finds the same constitutive relations
as those in Ref.~\cite{Hohenberg}:
\begin{eqnarray}
    t^{i j} &=& p\, \delta^{i j} + m (n-n_s) v^i v^j + m n_s v_s^i v_s^j , \\
    j_{\varepsilon}^i &=& n_s (\mu + m \mathbf{v} \cdot \mathbf{v}_s) 
                       (v_s^i - v^i) + (\varepsilon +p) v^i , \\ 
    \lambda^i &=& n_s(v_s^i - v^i). 
\end{eqnarray} 
 Note that in this case, the particle
current flow $j^i$ is itself conserved.
    
    
\section{Sum rules and hydrodynamic fluctuations}
\label{sec:fluc}
Here we will focus on the dynamics of fluctuations around 
the equilibrium state as seen from the rest frame of the superfluid. 
The velocity $\mathbf{v}(\mathbf{x},t)$ and $\bm{\nabla}\varphi$ will
be non-zero only due to a departure from the equilibrium state.
We apply the memory function formalism, quickly reviewed below, 
by following the treatment of Forster~\cite{Forster} for ${}^4$He.  
The linearized hydrodynamic equations including dissipative effects 
will be derived as a partial check of the previous results.  
  
The fundamental quantity in this discussion is the matrix of complex 
response functions
\begin{equation}
   \chi_{A B}(\mathbf{k},z)= \int_{-\infty}^\infty \frac{d\omega}{\pi} 
   \frac{\rho_{A B}(\mathbf{k},\omega)}{\omega-z} ,
\end{equation} 
which is analytic for $\mathrm{Im}\, z \neq 0$. 
For $z$ in the upper half-plane, $\chi_{A B}(\mathbf{k},\omega + i \epsilon)$ 
is the Fourier transform of 
the equilibrium expectation value of the retarded commutator 
$i \theta(t)\langle \left[A(\mathbf{x},t),B(\mathbf{0},0) \right]\rangle$,
and the spectral function $\rho_{A B}(\mathbf{k},\omega)$ is the Fourier transform 
of $\langle \left[A(\mathbf{x},t),B(\mathbf{0},0) \right]\rangle/2$. 
This quantity is either real and symmetric in $A\leftrightarrow B$, or
imaginary and antisymmetric.
If the operators $A$ and $B$ have the same
(opposite) signature under time reversal, $\rho_{A B}(\mathbf{k},
\omega)$ is odd (even) in $\omega$.  Such a signature is $+1$ for the
energy and charge densities and $-1$ for the momentum densities and
the Goldstone mode. 

When the system is perturbed by turning on a time-dependent Hamiltonian  depending on 
some set of small external forces coupled to the operators $\{A(\mathbf{x},t)\}$ 
\begin{equation}\label{eq:ext}
    \delta H^{\mathrm{ex}}(t) = -\sum_A \int d^3 \mathbf{x}\, A(\mathbf{x},t) 
    \delta F_A(\mathbf{x},t) \;, 
\end{equation}
the standard techniques of first-order perturbation theory produce 
the induced changes $\delta\langle A(\mathbf{x},t) \rangle$. 
In order to follow the relaxation of the induced quantities 
it is convenient to apply external fields that are held constant for 
negative times and are suddenly switch off for positive times 
\begin{equation}
   \delta F_A(\mathbf{x},t) = \delta F_A(\mathbf{x}) e^{\epsilon t} \theta(-t)\, ,
\end{equation}
where $\epsilon$ is a positive infinitesimal number. After elimination 
of the external fields in favour of the static susceptibilities, 
$\chi_{A B}(\mathbf{k}) = \chi_{A B}(\mathbf{k}, z=i \epsilon)$, through  
\begin{equation}\label{eq:forces}
\delta\langle A(\mathbf{k},t=0) \rangle = 
\sum_B \chi_{A B}(\mathbf{k})\, \delta F_B(\mathbf{k})\; ,
\end{equation}
the Laplace-Fourier transform of the 
induced changes $\delta\langle A(\mathbf{x},t) \rangle$ becomes~\cite{Forster} 
\begin{eqnarray}
\delta\langle A(\mathbf{k},z) \rangle &=& \sum_{B C}
\frac{1}{i z} \left(\chi_{A B}(\mathbf{k},z) \chi_{B C}^{-1}(\mathbf{k}) - 
\delta_{A C} \right)\delta\langle C(\mathbf{k},t=0) \rangle \nonumber \\ 
&\equiv& \beta \sum_{B C} C_{A B}(\mathbf{k},z) \chi_{B C}^{-1}(\mathbf{k})
\delta\langle C(\mathbf{k},t=0)\rangle \;.
\end{eqnarray}
This a fundamental result that solves the initial value problem in terms of 
response functions. 

General arguments based on the memory function formalism 
make it possible to write
the matrix $C$  in the form 
\begin{eqnarray}\label{eq:CAB}
C_{A B}(\mathbf{k},z) &=& \beta^{-1} \int_{-\infty}^\infty \frac{d\omega}{\pi i} 
   \frac{\rho_{A B}(\mathbf{k},\omega)}{\omega(\omega-z) } \nonumber \\  
   &=& \frac{i \beta^{-1}}{z - \omega(\mathbf{k})\chi^{-1}(\mathbf{k}) +
         i \sigma(\mathbf{k},z) \chi^{-1}(\mathbf{k})} 
         \,\chi(\mathbf{k})\;,  
\end{eqnarray}
where the matrices $\omega$ and $\sigma$ are given by 
\begin{eqnarray}
   \omega_{A B}(\mathbf{k}) &=& \int_{-\infty}^\infty\frac{d\omega}{\pi} 
   \rho_{A B}(\mathbf{k},\omega)\; , \\ 
   \sigma_{A B}(\mathbf{k},z) &=& \int_{-\infty}^\infty\frac{d\omega}{2 \pi i} 
   \frac{\gamma_{A B}(\mathbf{k},\omega)}{\omega -z} \; ,
\end{eqnarray}
and they 
encode reactive and relaxations properties of the system, respectively. 
The spectral density  $\gamma_{A B}(\mathbf{k},\omega)$ has the 
same symmetry properties as 
those of the matrix $\omega \rho_{A B}(\mathbf{k},\omega)$ 
and must define a positive quadratic form~\cite{Forster} from the 
requirement of positive entropy production.  
For $z=0$, $\sigma_{AB}(\mathbf{k},z)$ corresponds to a matrix of
transport coefficients.  Thus, in this framework, the input of
$\chi_{A B}(\mathbf{k})$, $\sigma_{A B}(\mathbf{k},z=0)$ and
$\omega_{A B}(\mathbf{k})$ for small $k$ is all that is required in
order to extract hydrodynamic correlation functions.   
With these definitions, the linearized hydrodynamic equations adopt the form
\begin{equation}\label{eq:linhydro}
 z \delta \langle A(\mathbf{k},z) \rangle - 
 \sum_{B}\left[\omega_{AB}(\mathbf{k}) - i \sigma_{AB}(\mathbf{k},0)\right] 
 \delta F_B(\mathbf{k},z) = i \delta\langle A(\mathbf{k},t=0)\rangle \;, 
\end{equation}
where $\{\delta F(\mathbf{k},z)\}$ is the set of internal forces 
expressing the departures of the thermodynamic quantities 
from its equilibrium values. 
By analogy with Eq.~(\ref{eq:forces})
they are defined for $t>0$ by  
\begin{equation}\label{eq:internal}
\delta F_A(\mathbf{k},t) = \sum_{B}\chi_{A B}^{-1}(\mathbf{k}) 
\delta \langle B(\mathbf{k},t)\rangle.
\end{equation}
  
To proceed further, we need some sum rules determining the 
appropriate $\chi(\mathbf{k})$ and $\omega(\mathbf{k})$. 
These have been derived in Ref.~\cite{Baym} and, essentially,  
they do not differ from the corresponding ones in 
nonrelativistic superfluids~\cite{Forster}.
We present the list of the required equations.  

The split of the momentum density $\mathbf{g}$ into irrotational  
$\mathbf{g}_{\mathrm{L}}$ and solenoidal $\mathbf{g}_{\mathrm{T}}$ parts
leads to the separation of the momemtum density response function  
$\chi_{g g}^{i j}(\mathbf{k},\omega) = 
\chi_{\mathrm L}(k,\omega)\hat{k}^i\hat{k}^i + \chi_{\mathrm T}(k,\omega)(\delta^{i j}-
\hat{k}^i\hat{k}^i)$ into longitudinal and transverse pieces.
For a normal relativistic fluid, the momentum density is given
by $\mathbf{g} = h \mathbf{v}$ and  
the  momentum susceptibility is $\partial g^i/\partial v^j = \delta^{i j} h$
where $h$ is the enthalpy density.
In the superfluid phase the momentum density acquires an extra
irrotational contribution $\mathbf{g}_s \propto \bm{\nabla} \varphi$
\begin{equation}\label{eq:grel}
\mathbf{g} = (h - h_s) \mathbf{v}+\mathbf{g}_s , 
\end{equation}
where $h_s$ is the superfluid enthalpy density. 
The superfluid momentum density 
arises from long-range order due to the Goldstone mode whose
correlation function behaves as 
\begin{equation}\label{eq:fifi}
   \chi_{\varphi \varphi}(k)=\int_{-\infty}^\infty \frac{d\omega}{\pi} 
          \frac{\rho_{\varphi \varphi}(k,\omega)}{\omega}= \frac{1}{\xi k^2}
          \quad 
   \textrm{as $k\rightarrow 0$},   
   \end{equation} 
where the constant $\xi$ is positive. 
Thus the momentum susceptibility has a normal (isotropic) 
contribution $(h-h_s) \delta^{i j}$ 
and a superfluid contribution $h_s \hat{k}^i\hat{k}^i$ as $k \rightarrow 0$.
From the arrangement $(h-h_s) \delta^{i j} + h_s \hat{k}^i\hat{k}^j = 
h \hat{k}^i\hat{k}^j+(h-h_s) (\delta^{i j}-\hat{k}^i\hat{k}^j)$
we can write the following sum rules 
\begin{eqnarray}
   \label{eq:chiL}
   \lim_{k \rightarrow 0}\chi_{\mathrm L}(k) & = & \lim_{k \rightarrow 0} \int_{-\infty}^\infty \frac{d\omega}{\pi} 
          \frac{\rho_{\mathrm{L}}(k,\omega)}{\omega}= h  \; , \\
   \lim_{k \rightarrow 0}\chi_{\mathrm{T}}(k) & = & \lim_{k \rightarrow 0}\int_{-\infty}^\infty \frac{d\omega}{\pi} 
          \frac{\rho_{\mathrm{T}}(k,\omega)}{\omega} = h-h_s \; . 
\end{eqnarray}

Other sum rules involving $\varphi$ are
\begin{eqnarray}\label{eq:epshi}
\omega_{\varepsilon \varphi}(k)=
\int_{-\infty}^\infty \frac{d\omega}{\pi} 
          \rho_{\varepsilon \varphi}(k,\omega)&=& i \mu  \;, \\ 
\label{eq:chigpshi}
   \chi_{g \varphi}^{i}(\mathbf{k}) =\int_{-\infty}^\infty \frac{d\omega}{\pi} 
          \frac{\rho_{g \varphi}^j(\mathbf{k},\omega)}{\omega}&=& \frac{i \mu k^j}{k^2}\; ,
\end{eqnarray}
where Eq.~(\ref{eq:epshi}) is a consequence of the averaged Heisenberg equation of motion
\begin{equation}
    -i \partial_t \langle \varphi(\mathbf{x},t)\rangle = \langle
\left[H, \varphi(\mathbf{x},t) \right] \rangle = i \mu ,
\end{equation}
and $\chi_{g \varphi}^{i}(\mathbf{k})$ follows  
from energy conservation.
The derivation of Eq.~(\ref{eq:epshi}) can be made more rigorous by directly
averaging on a restricted $\eta$-ensemble~\cite{Hohenberg} appropriate
to superfluids\footnote{Details of a similar computation can be found
in p.  237 of Ref.~\cite{Forster}.}.  These results together with
Eq.~(\ref{eq:fifi}) lead to the proportionality constant between
$\mathbf{g}_s$ and $\bm{\nabla}\varphi$.  As $\chi_{g_s g_s}^{i j} =
\chi_{g g_s}^{i j}= h_s \hat{k}^i\hat{k}^j$, consistency requires that
\begin{equation}\label{eq:gs2}
      \mathbf{g}_s= \xi \mu  \bm{\nabla}\varphi,  \quad h_s = \xi \mu^2 ,
\end{equation}
in precise agreement with the linealized expression of the momentum
density in Eq.~(\ref{eq:gs}).  Note that the quantity $\mathbf{v}_s =
\mu^{-1} \bm{\nabla}\varphi$ plays the role of the conventional
superfluid velocity.

On the other hand,
Eq.~(\ref{eq:chiL})
and energy conservation yield
   \begin{equation}
   \int_{-\infty}^\infty \frac{d\omega}{\pi} 
          \rho_{\varepsilon g}^{\;\; j}(\mathbf{k},\omega) = h k^j  
   \end{equation}
omitting terms of higher order in $k$.
Also, the equal time conmutators 
\begin{eqnarray}
   \left[n(\mathbf{x},t), \mathbf{g}(\mathbf{y},t) \right]&=& 
    i\, n(\mathbf{y},t) \bm{\nabla}_y\delta(\mathbf{x}-\mathbf{y})  , \\ 
   \left[n(\mathbf{x},t), \varphi(\mathbf{y},t) \right]&=& 
    i \delta(\mathbf{x}-\mathbf{y}) , 
\end{eqnarray}
produce  
\begin{eqnarray} 
   \int_{-\infty}^\infty \frac{d\omega}{\pi} 
          \rho_{n g}^{\;\; j}(\mathbf{k},\omega)&=& n k^j  \;, \\ 
   \int_{-\infty}^\infty \frac{d\omega}{\pi} 
          \rho_{n \varphi}(k,\omega)&=& i \, .        
\end{eqnarray}
   
All these sum rules are valid irrespective of whether the system with 
broken symmetry is relativistic or not. 
In the nonrelativistic case where the mass density is not
included in the energy density, the right side of Eq.~(\ref{eq:epshi}) 
must be replaced by $i \mu_{\mathrm{NR}}$, 
according to the usual definition $\mu=m +
\mu_{\mathrm{NR}}$.
Note, however, that in 
Eq.~(\ref{eq:chigpshi}) the chemical potential must be replaced by its 
leading part $m$. 
This gives the correct
values appropriate to the nonrelativistic superfluid, $h_s = m n_s$,
$h=m n$ and $\xi = n_s/m$, where $n_s$ is the superfluid particle
density.


\section{Correlation functions and Kubo relations} 
\label{sec:kubo} 
With these results in hand, and the definition 
$g_{\mathrm L}(\mathbf{k},t)=
\mathbf{g}(\mathbf{k},t)\cdot \hat{\mathbf{k}}$, 
we can write the matrices $\chi$ and $\omega$
corresponding to the four hydrodynamic variables 
$\{\varepsilon, g_{\mathrm L}, n, \varphi\}$
but the subsequent formulae will be considerably simplified if 
in place of the particle density $n$
we introduce a new variable $q$ defined by~\cite{Kadanoff}
\begin{equation}
dq = T n d\left(\frac{s}{n}\right)=d\varepsilon - \frac{h}{n}\, dn .  
\end{equation}
In the linearized theory this quantity corresponds  to the combination 
$\varepsilon(\mathbf{k},t) -
(h/n)_{\mathrm{eq}} n(\mathbf{k},t)$ representing
the density of heat energy.  
The simplification arises  
because the term $\omega_{q g_{\mathrm L}}$ becomes 
zero,  $\omega_{\varepsilon g_{\mathrm L}} - 
 h/n \omega_{n g_{\mathrm L}}=0$, and the only non zero $\omega$-terms  involving $q$ are 
 $\omega_{q \varphi} = -\omega_{\varphi q}=-i s T/n$ as follows from the Gibbs-Duhem identity. 
Therefore, in terms of the hydrodynamic variables
\begin{equation}
      \{O_A(\mathbf{k},t) \} = \{\varepsilon, g_{\mathrm L}, 
      q, \varphi\}(\mathbf{k},t),  
\end{equation}
and two spectral functions 
\begin{eqnarray}
 \rho_{\varepsilon g_{\mathrm{L}}}(k,\omega) &\equiv& 
  \rho_{\varepsilon g}^{\;\; j}(\mathbf{k},\omega)\hat{k}^j, \\ 
 \rho_{g_{\mathrm{L}} \varphi}(k,\omega) &\equiv& 
  \rho_{g \varphi}^j(\mathbf{k},\omega)\hat{k}^j, 
\end{eqnarray}
real and symmetric, imaginary and antisymmetric respectively, 
the required matrices $\chi$ and $\omega$ are given by 
\begin{equation}\label{eq:susce}
   \chi_{A B}(k) = \left(\begin{array}{cccc} 
     \chi_{\varepsilon \varepsilon} & 0 & \chi_{\varepsilon q} & 0 \\ 
     0   & h & 0 & \frac{i \mu}{k} \\ 
     \chi_{\varepsilon q} & 0 & \chi_{q q} & 0  \\ 
     0 & -\frac{i \mu}{k} & 0 & \frac{1}{\xi k^2} 
                      \end{array} \right) , 
\end{equation}
and 
\begin{equation}
   \omega_{A B}(k) = \left(\begin{array}{cccc} 
     0 & h k & 0 & i \mu \\ 
     h k   & 0 & 0 & 0 \\ 
     0 & 0 & 0 & -i \frac{T s}{n}  \\ 
     -i \mu  & 0  & i \frac{T s}{n} & 0 
                      \end{array} \right) , 
\end{equation}
where the remainder vanishing matrix elements are due to time reversal symmetry. 
The susceptibilities may be obtained from the thermodynamic potential 
$\Omega = -V p(T,\mu)$ by differentiation. 
These matrices are completed with the results for the transverse momentum density,
  \begin{eqnarray}
   \chi_{g_{\mathrm T} g_{\mathrm T}}^{i j} &=& (h-h_s) (\delta^{i j} - \hat{k}^i \hat{k}^j), \\ 
   \omega_{g_{\mathrm T} g_{\mathrm T}}^{i j} &=& 0. 
  \end{eqnarray}

From the thermodynamical derivatives
\begin{eqnarray}
\left(\frac{\partial p}{\partial \varepsilon}\right)_{s/n}&=& 
\frac{h \chi_{qq}}{\chi_{\varepsilon \varepsilon}\chi_{qq} - 
                  \chi_{\varepsilon q}^2} , \\ 
\left(\frac{\partial p}{\partial q}\right)_{\varepsilon}&=& 
\frac{-h \chi_{\varepsilon q}}{\chi_{\varepsilon \varepsilon}\chi_{qq} - 
                  \chi_{\varepsilon q}^2} , \\
\left(\frac{\partial (\mu/T)}{\partial \varepsilon}\right)_{s/n}&=&
\frac{h}{n T} \frac{\chi_{\varepsilon q}}{\chi_{\varepsilon \varepsilon}\chi_{qq} - 
                  \chi_{\varepsilon q}^2} ,  \\ 
\left(\frac{\partial (\mu/T)}{\partial q}\right)_{\varepsilon}&=&
-\frac{h}{n T} \frac{\chi_{\varepsilon \varepsilon}}{\chi_{\varepsilon \varepsilon}\chi_{qq} - 
                  \chi_{\varepsilon q}^2} ,                                 
\end{eqnarray}
and Eqs.~(\ref{eq:grel}), (\ref{eq:gs2}) and~(\ref{eq:susce}) 
we obtain the internal forces easily
\begin{eqnarray} 
\delta F_\varepsilon  &=& \frac{\delta p}{h} , \\
\delta \mathbf{F}_{g}&=& \mathbf{v} , \\
\delta F_q  &=&  -\frac{n T}{h}\, \delta\left(\frac{\mu}{T} \right) , \\
\delta \mathbf{F}_{\varphi}&=& 
    \xi (-\nabla^2 \varphi + \mu \bm{\nabla}\cdot \mathbf{v}).
\end{eqnarray}   
Note that these last results depend on the choice of the linearized hydrodynamical variables. 
So, in terms of $\{q, g_{\mathrm L}, n, \varphi\}$ the internal forces conjugate to energy density and 
particle density are $\delta F_\varepsilon =T^{-1} \delta T$ and 
$\delta F_n  =T\delta(\mu/T)$. 

Now, we look at the memory matrix $\sigma_{AB}(\mathbf{k},0)$ to lowest order in $k$. 
The transverse memory function reads 
  \begin{equation}
  \sigma_{g_{\mathrm T} g_{\mathrm T}}^{i j}(\mathbf{k},0) = 
    \eta k^2 (\delta^{i j} -\hat{k}^i \hat{k}^j),
  \end{equation} 
where $\eta > 0$ is the shear viscosity. With the previous notation, 
  the remainder elements can be parametrized by 
  \begin{equation}
   \sigma_{A B}(\mathbf{k},0) = \left(\begin{array}{cccc} 
     0 & 0 & 0 & 0 \\ 
     0 & \left(\zeta_2 + \frac{4}{3} \eta \right) k^2 & 0 & i \zeta_1 k \\ 
     0 & 0 & \kappa T k^2 & 0  \\ 
     0 & -i \zeta_1 k & 0 & \zeta_3 
    \end{array} \right) ,  
   \end{equation}
in terms of the thermal conductivity $\kappa$ and three 
non negative longitudinal viscosities 
  $\zeta_1$, $\zeta_2$, $\zeta_3$  
with the constraint
  \begin{equation}
      \zeta_2 \zeta_3 \geq \zeta_1^2 ,
  \end{equation}
  from the positivity of $\sigma$.  
  The $\sigma$-matrix has been chosen with the 
  requirement that all elements involving the energy density vanish 
  at leading order in $k$. 
  This is the main difference with the corresponding matrix for 
  the nonrelativistic superfluid in terms of variables
  $\{q, g_{\mathrm L}, n, \varphi\}$
  \begin{equation}
   \sigma_{A B}^{\mathrm{NR}}(\mathbf{k},0) = \left(\begin{array}{cccc} 
     \kappa T k^2 & 0 & 0 & 0 \\ 
     0 & \left(\zeta_2 + \frac{4}{3} \eta \right) k^2 & 0 & i \zeta_1 k \\ 
     0 & 0 & 0 & 0  \\ 
     0 & -i \zeta_1 k & 0 & \zeta_3 
    \end{array} \right) ,  
  \end{equation}
  where the vanishing elements are those involving the particle density $n$. 
  The reason for these choices lies in the fact that when a 
  current flow is itself a conserved quantity, 
  the dissipative lowest order contribution to the constitutive relation 
  vanishes~\cite{Forster}. 
It is worthwhile to remark that the absence of transport coefficients for  
the energy current flow, $j^i_\varepsilon = g^i$,  is only consistent with the usual relativistic 
dissipative fluid theory of Landau and Lifshitz and not with 
that of Eckart. 
  
All the hydrodynamic correlation functions 
$\chi_{A B}(\mathbf{k},\omega)$ 
can be now derived from Eq.~(\ref{eq:CAB}) 
with similar results to 
those reported in Ref.~\cite{Hohenberg} for the nonrelativistic case. 
They have poles at the lower half-plane when their denominator
\begin{equation}
\Delta(k,\omega) = (\omega^2 - c_1^2 k^2 + i k^2 D_1 \omega)(\omega^2 - c_2^2 k^2 + i k^2 D_2 \omega)
\end{equation}
vahishes, except for the transverse correlation function with a 
diffusive pole at $\omega = -i \eta k^2/(h-h_s)$.
In terms of the square of the velocity of adiabatic sound waves
\begin{equation}
c^2 = \left(\frac{\partial p}{\partial \varepsilon}\right)_{s/n}=
\frac{h \chi_{qq}}{\chi_{\varepsilon \varepsilon}\chi_{qq} - 
                  \chi_{\varepsilon q}^2}, 
\end{equation}
and the normal enthalpy density $h_{\mathrm{n}} =h-\xi \mu^2$,
the speeds of propagation of first and second sound are given by 
\begin{eqnarray}
c_1^2+c_2^2 &=& c^2 \left(1+ \frac{\xi\, T^2 s^2 \chi_{\varepsilon \varepsilon}}
                  {n^2 h_{\mathrm{n}} \chi_{qq}}  \right), \\
c_1^2 c_2^2 &=&\frac{c^2 \xi\, h\, T^2 s^2}
                  {n^2 h_{\mathrm{n}} \chi_{qq}} ,
\end{eqnarray}  
and the attenuation constants can be written in the form
\begin{eqnarray}
D_1+D_2 &=& \frac{1}{h_{\mathrm{n}}}\left(\zeta_2+\frac{4}{3}\eta\right)+
            \frac{\xi (h \zeta_3 - 2 \mu \zeta_1)}{h_{\mathrm{n}}} + 
            \frac{c^2 T \chi_{\varepsilon \varepsilon}\kappa}{h \chi_{qq}} , \\
c_1^2 D_2+c_2^2 D_1 &=& \frac{c^2 \xi 
                (n^2 \mu^2 \chi_{qq} + 2 n s T \mu \chi_{\varepsilon q}+
                s^2 T^2 \chi_{\varepsilon \varepsilon})}{n^2 h h_{\mathrm{n}}\chi_{qq}}
                \left(\zeta_2+\frac{4}{3}\eta\right)  \nonumber \\ 
                &&+ \frac{c^2 \xi h \zeta_3}{h_{\mathrm{n}}} - 
                    \frac{2 c^2 \xi (n \mu\chi_{qq} + s T \chi_{\varepsilon q}) \zeta_1}
                    {n h_{\mathrm{n}}\chi_{qq}} \nonumber \\ 
                &&+ \frac{c^2 T \kappa}{\chi_{qq}}.     
\end{eqnarray}  
Some Ward identities can be directly checked. We list them 
\begin{eqnarray}
\omega \chi_{\varepsilon \varepsilon}(k, \omega)&=&
    k \chi_{g_{\mathrm L} \varepsilon}(k, \omega), \\
\omega \chi_{\varepsilon g_{\mathrm L}}(k, \omega)&=&
    k \chi_{\mathrm L}(k, \omega) - h k, \\
\omega \chi_{\varepsilon n}(k, \omega)&=&
    k \chi_{g_{\mathrm L} n}(k, \omega), \\
\omega \chi_{\varepsilon \varphi}(k, \omega)&=&
    k \chi_{g_{\mathrm L}\varphi}(k, \omega) - i \mu . 
\end{eqnarray}
However, 
Ward identities for correlation functions involving 
currents such as $t^{k l}$ or $j^i$ cannot be checked 
within this approach because these flows are not hydrodynamical modes.

The usual catalog of Kubo relations giving the transport coefficients 
can be obtained from the following limits
\begin{eqnarray}
\eta &=& \lim_{\omega \rightarrow 0}\lim_{k \rightarrow 0} 
      \frac{\omega}{k^2}\,\mathrm{Im}\,\chi_{\mathrm{T}}(k,\omega) , \\
\zeta_2 + \frac{4}{3} \eta  &=& 
  \lim_{\omega \rightarrow 0}\lim_{k \rightarrow 0} 
      \frac{\omega}{k^2}\,\mathrm{Im}\,\chi_{\mathrm{L}}(k,\omega) , \\      
\kappa T \left(\frac{n}{h}\right)^2 &=& \lim_{\omega \rightarrow 0}\lim_{k \rightarrow 0} 
      \frac{\omega}{k^2}\,\mathrm{Im}\,\chi_{n n}(k,\omega) , \\
 \zeta_1 &=& -\lim_{\omega \rightarrow 0}\lim_{k \rightarrow 0} 
      \frac{\omega}{k}\,\mathrm{Re}\,\chi_{g_\mathrm{L}\,\varphi}(k,\omega) , \\ 
 \zeta_3 &=& \lim_{\omega \rightarrow 0}\lim_{k \rightarrow 0} 
      \omega \,\mathrm{Im}\,\chi_{\varphi \,\varphi}(k,\omega) ,          
\end{eqnarray}
where for the thermal conductivity we have replaced $\chi_{qq}$ by 
$(h/n)^2 \chi_{nn}$ 
since
\begin{equation}
\lim_{\omega \rightarrow 0}\lim_{k \rightarrow 0} 
      \frac{\omega}{k^2}\,\mathrm{Im}\,\chi_{\varepsilon \varepsilon}(k,\omega) =
\lim_{\omega \rightarrow 0}\lim_{k \rightarrow 0} 
      \frac{\omega}{k^2}\,\mathrm{Im}\,\chi_{\varepsilon q}(k,\omega)=0. 
\end{equation}
The shear and bulk viscosities $\eta$ and $\zeta_2$ 
of the superfluid quark matter in the color-flavor locked phase have been computed in 
Refs.~\cite{Manuel1,Manuel2}, and all the bulk viscosities in neutron stars 
have been computed in Ref.~\cite{Gusakov}.

Finally, the constitutive relations from Eq.~(\ref{eq:linhydro}) 
read
\begin{eqnarray}
\delta \langle \mathbf{j}_\varepsilon\rangle &=&
(h - \xi \mu^2) \mathbf{v} + \xi \mu \bm{\nabla} \varphi , \\
\delta \langle \mathbf{j}_n\rangle & =& (n -\xi \mu) \mathbf{v} + 
                 \xi \bm{\nabla} \varphi
              -\kappa \left(\frac{n T}{h}\right)^2
                  \bm{\nabla} \left(\frac{\mu}{T}\right) , \\
\delta \mu^{\mathrm{total}}(\mathbf{x},t) &=& \delta \mu(\mathbf{x},t) - 
  \zeta_1 \bm{\nabla} \cdot \mathbf{v} -
  \xi \,\zeta_3\,(\nabla^2 \varphi - \mu \bm{\nabla}\cdot \mathbf{v}) , \\ 
\delta \langle t^{i j}\rangle &=& \delta p(\mathbf{x},t) \delta^{i j} - 
      \eta \left(\nabla_i v_j + \nabla_j v_i - 
              \frac{2}{3} \bm{\nabla}\cdot\mathbf{v}\, \delta^{i j}\right) \nonumber \\ 
              && - 
              \delta^{i j}\left(\zeta_2 \bm{\nabla} \cdot \mathbf{v} + 
                 \xi \zeta_1 (\nabla^2 \varphi - \mu \bm{\nabla}\cdot \mathbf{v}) \right)  .
\end{eqnarray}
The reactive parts of these equations agree to linear order with those found in 
Sec.~\ref{sec:bra}. 

Summarizing, we have presented the distinctive features of relativistic superfluids 
(in comparison with those in the nonrelativistic regime) accounting for their 
hydrodynamics
and the  
correlation functions have been derived. The crucial assumption is 
that the momentum density coincides with the flow of the conserved energy. 
As a consequence, there are no 
dissipative contribution to the energy current and the thermal conductivity appears in  
the current of the conserved particle number, 
according to the fluid theory of  Landau and Lifshitz.
  


\begin{acknowledgments}
I thank I\~nigo Egusquiza, Juan L. Ma\~nes and Christian Romelsberger for useful discussions.
This work has been supported in part by the Spanish Ministry of Science and Technology 
under grant FPA2005-04823. 
\end{acknowledgments}


\end{document}